\documentclass[twocolumn]{aastex631}

\submitjournal{ApJ}
\accepted{November 3, 2021}
\shortauthors{M. De Furio et al.}
\shorttitle{Brown Dwarf Binaries in the ONC}
\graphicspath{{./}{figures/}}

\begin{document}

\title{Binary Formation in the Orion Nebula Cluster: Exploring the Sub-stellar Limit}

\correspondingauthor{Matthew De Furio}
\email{defurio@umich.edu}

\author{Matthew De Furio}
\affiliation{Department of Astronomy, University of Michigan, Ann Arbor, MI}

\author{Michael R. Meyer}
\affiliation{Department of Astronomy, University of Michigan, Ann Arbor, MI}

\author{Megan Reiter}
\affiliation{European Southern Observatory}

\author{John Monnier}
\affiliation{Department of Astronomy, University of Michigan, Ann Arbor, MI}

\author{Adam Kraus}
\affiliation{Department of Astronomy, University of Texas - Austin, Austin, TX}


\author{Trent Dupuy}
\affiliation{University of Edinburgh, Edinburgh, United Kingdom}



\begin{abstract}

We present results constraining the multiplicity of the very low mass stars and sub-stellar objects in the Orion Nebula Cluster (ONC).  Our sample covers primary masses 0.012-0.1M$_{\odot}$ using archival Hubble Space Telescope data obtained with the Advanced Camera for Surveys using multiple filters.  Studying the binary populations of clusters provides valuable constraints of how the birth environment affects binary formation and evolution.  Prior surveys have shown that the binary populations of high-mass, high-density star clusters like the ONC may substantially differ from those in low-mass associations.  Very low mass stellar and sub-stellar binaries at wide separations, $>$20AU, are statistically rare in the Galactic field and have been identified in stellar associations like Taurus-Auriga and Ophiuchus.  They also may be susceptible to dynamical interactions, and their formation may be suppressed by feedback from on-going star formation.  We implement a double point-spread function (PSF) fitting algorithm using empirical, position dependent PSF models to search for binary companions  at projected separations $>$ 10 AU (0.025”).  With this technique, we identify 7 very low mass binaries, 5 of which are new detections, resulting in a binary frequency of 12$^{+6}_{-3.2}$\% over mass ratios of 0.5 - 1.0 and projected separations of 20 - 200 AU.  We find an excess of very low mass binaries in the ONC compared to the Galactic field, with a probability of 10$^{-6}$ that the populations are statistically consistent. The sub-stellar population of the ONC may require further dynamical processing of the lowest binding energy binaries to resemble the field population.
\end{abstract}



\section{Introduction} \label{sec:intro}
Developing a predictive theory of star formation requires an explanation for the formation of low-mass stars and sub-stellar objects ($<$ 0.1 M$_{\odot}$).  They dominate in number and represent a limiting case of star formation at the low mass end of the initial mass function (IMF).  Many star cluster studies are designed to constrain this low mass portion of the IMF where brown dwarfs account for roughly one object for every five stars with masses $<$ 1 \(\textup{M}_\odot\) \citep{Andersen2008}.  The multiplicity of such objects puts important constraints on specific companion formation pathways based on primary mass, e.g. turbulent core fragmentation \citep{Goodwin2004, Offner2010}, disk fragmentation \citep{Adams1989, Bonnell1994}, and others.  Since multiplicity is a common outcome of star formation, we must explore the low mass end of the companion population to identify any incongruities between stellar and sub-stellar primaries.

Stars and brown dwarfs tend to form in clusters or associations which will eventually dissolve into the Galactic field \citep{Carpenter2000}.  The initial population of binaries within young clusters precedes the eventual makeup of the field binary population.  Various processes, such as dynamical interactions and stellar feedback, can influence the formation and evolution the binary populations of clusters \citep{Kroupa2001}.  Such processes vary based on the environment of the clusters: higher stellar density increases the frequency of dynamical interactions \citep{Weinberg1987}, and stellar feedback from high mass stars may suppress low-mass star formation \citep{Krumholz2006,Liu_2017}.  We can study the primordial binary populations of diverse star forming regions to search for any differences relative to the Galactic field and as a function of environmental properties such as cluster mass and density.

Brown dwarf multiplicity studies in the Galactic field reveal companions at close separations, $<$ 10 AU, with a paucity of wide companions, $>$ 20 AU \citep{Close2003, Gizis2003, Reid2006}.  Although rare, wide companions to brown dwarfs have been found in the field \citep[e.g.][]{Billeres2005, Burningham2010, Dhital2011, Faherty2020}.

Other studies targeting young, stellar associations such as Upper Sco, Chamaeleon I, TW Hydrae, and Ophiuchus have also identified wide companions to brown dwarfs \citep[e.g.][]{Chauvin2004, Luhman2004, Close2007, Ahmic2007}.  These surveys typically have small sample sizes which leads to large uncertainties in binary frequency.  They often target nearby low-mass stellar associations which could contribute a smaller fraction of stars to the Galactic field than denser regions \citep{patience2002}, leaving open the need for both larger surveys and surveys in high density, high mass star forming regions.

The Orion Nebula Cluster (ONC) is the nearest \citep[400pc,][]{oriondistance}, young \citep[mean age of 2.2 Myr,][]{Reggiani2011} high mass star-forming region and an ideal location to study the influence of stellar density and cluster size on binary formation.  Previous multiplicity surveys of the ONC have mainly focused on companions to stellar primaries \citep{Reipurth2007,Duchene2018,DeFurio2019}.  Only recently have candidate companions to brown dwarfs been identified in the ONC \citep{Robberto2020,Strampelli2020}.  Using the IR channel of Wide
Field Camera 3 on the \textit{Hubble Space Telescope} (\textit{HST}), \citet{Strampelli2020} identified candidate companions down to separations of 0.16", $\sim$ 64 AU in the ONC.  This leaves an unexplored parameter space around brown dwarf primaries in the ONC at separations below $\sim$ 64 AU, where companions are more often found around field brown dwarfs.

The \textit{HST} Treasury Program on the ONC \citep{Robberto2013} using the Advanced Camera for Surveys (ACS) has surveyed the large area of the ONC, providing high resolution imaging with a diffraction limit of $\sim$ 0.05" in the F435W filter to $\sim$ 0.09" in the F850LP filter.  The stable point-spread function (PSF) of space-based imaging permits the use of double PSF-fitting with empirical models to attain high spatial resolution to detect companions near the diffraction limit.  Building on the procedure of \citet{DeFurio2019}, hereafter Paper I, we use double-PSF fitting to identify companions at separations down to $\sim$ 10 AU (0.025" at 400 pc), rivaling the resolution of aperture-masking interferometry with adaptive optics on 8m telescopes \citep[e.g.][]{Duchene2018} and kernel phase interferometry with HST \citep[e.g.][]{Pope2013} over wide fields of view, opening up an interval of separations completely unexplored for brown dwarfs in the ONC.

In Section 2, we describe the ONC sample, our double-PSF model, and the improvements to the method introduced in Paper I.  In Section 3, we present the newly discovered binaries, describe our detection limits, and define the companion population for brown dwarfs in the ONC.  In Section 4, we compare the ONC to the Galactic field and various star clusters and associations, and we discuss the implications of our results.  In Section 5, we summarize our conclusions.


\section{Methods} \label{sec:methods}
The double-PSF fitting code that we apply to the imaging data in this paper is identical to that of Paper I.  Where our analysis differs is in the examination of the outputs from the double-PSF algorithm as well as a more rigorous investigation of the completeness and reliability of the code.

\subsection{The Data} \label{subsec:data}
We downloaded multi-filter data from the Treasury Program of the Orion Nebula Cluster (GO program 10246, PI: M. Robberto) from the \textit{HST} archive.  Our analysis utilizes data taken in the F435W, F555W, F775W, and F850LP filters with the Advanced Camera for Surveys in the Wide Field Channel mode (ACS/WFC) over a series of 104 orbits during Cycle 13.  All data were taken over the course of a six month time period.  The integration time for each exposure was 420s in the F435W filter and 385s in the F555W, F775W, and F850LP filters.  The Treasury Program survey covered $\sim$ 600 square arcminutes where most of the imaged area has at least two exposures.  We use the data products with the suffix $\_$flt, which have been bias-subtracted and flat-fielded,  and are the images to which we applied the empirical PSFs of \citet{AndersonKing2006}.  ACS/WFC has a plate scale of 0.05"/pixel, where the PSF full-width half maximum is 0.080", 0.088", 0.077", and 0.088" for the F435W, F555W, F775W, and F850LP filters respectively \citep{Clampin2003, Windhorst2011}.  The PSFs of the ACS/WFC detector are under-sampled, and PSF-fitting with empirical models is an accurate way to determine the centroid of the PSF and identify close companions.  A detailed explanation of observations and data reduction can be found in \citet{Robberto2013}.

\subsection{The Model} \label{subsec:model}
We described the binary modelling in Paper I, but the following is a short summary.  For each image of any given source, we use a 21x21 pixel postage stamp centered on the target as the input to our algorithm.  Before creating a postage stamp, we subtract the mean background centered around the source in an annulus of inner and outer radii of 10 and 15 pixels, respectively.  Then, we fit a binary star model consisting of two individual PSFs to each postage stamp.

The PSFs are constructed from the 4x super-sampled, empirically derived PSF model libraries of \citet{AndersonKing2006}, hereafter AK06, created in multiple filters for ACS/WFC, which describe the structure of the PSF as a function of position on the ACS detectors.  Each filter listed in Sec. 2.1 has its own two dimensional PSF library generated by AK06 except for F555W.  Instead, we use the F606W PSF library as the differences in PSF structure between these two filters is negligible (J. Anderson, private communication).  Additionally, AK06 include a PSF perturbation function which alters the shape of the empirical-PSF model libraries based on the PSFs of the brightest sources in an individual data image.  This procedure mitigates the effects of changes in focus and instabilities in pointing which can cause the PSF to deviate from the average over time.  We apply this perturbation to the PSF libraries of each filter in each image with ONC targets.

We apply a bi-cubic interpolation to these super-sampled, empirical PSFs, as done in \citet{AndersonKing2000,AndersonKing2003,AndersonKing2004,AndersonKing2006}, in order to construct a detector-sampled PSF.  We perform this action twice for a given data stamp to fit two detector-sampled, empirical PSFs.  Our double-PSF model of a binary system that performs this interpolation routine has six parameters: x and y coordinate position of the primary, separation between the centers of the primary and companion PSFs, position angle corresponding to the location of the center of the companion, the overall magnitude of the system, and the difference in magnitude between the primary and companion ($\Delta$mag).  The x and y coordinates dictate where the first detector-sampled PSF will be created within the data stamp.  The separation and position angle parameters dictate where the second detector-sampled PSF will be created within in the data stamp.  The combined magnitude and $\Delta$mag of the system will determine the brightness of these detector-sampled PSFs within each postage stamp.

We initialize the double PSF-fitting routine with the results of a single-PSF fit to each individual source. We use the x and y position of the primary and the combined magnitude of the system from the single-PSF fit as the starting values for a coarse grid search.  Then, we perform the coarse grid search in the six dimensional parameter space of the binary model, calculating the chi squared test statistic for each fit.  We take the best fit from the coarse grid search, add small random offsets, and use the result as the starting position for a downhill simplex algorithm (AMOEBA) in order to avoid falling into a local minimum.  The AMOEBA routine steps in the six dimensional parameter space according to a series of expansions, contractions, and reflections based on the corresponding $\chi^{2}_{\nu}$ value of each fit before converging after achieving some fractional tolerance between iterations.  AMOEBA is run 200 times from a slightly different starting position due to the random perturbations.  We find the best fit (lowest chi squared test statistic, as described in Paper I) binary star model from the 200 runs to any input data with errors, regardless of whether the system is a binary.  This process is performed individually for each image of all targets across all filters in which the targets appear.  To determine if a given double-PSF fit is a true detection of a binary system, we calculate the false detection probability for each source in each image of each filter based on our sensitivity tests and metrics as stated in Sec. \ref{subsec:falsepositives}.

\subsection{The Sample} \label{subsec:subsample}
We adopted our input target list of M6 - M9.5 objects ($\sim$ 0.012 - 0.1M$_{\odot}$ at 2 Myr from \citet{Allard2001} evolutionary models) from \citet{Hillenbrand2013} who identified ONC members with medium resolution, red optical spectroscopy. They classified low mass objects by measuring the depth of TiO and VO lines that are mainly temperature sensitive \citep{Riddick2007}. We also included M6 - M9.5 members that were previously spectroscopically confirmed in other surveys \citep[e.g.][]{Lucas2001,Slesnick2004,Lucas2006, Riddick2007, Weights2009} with optical spectra using TiO and VO band indices or near infrared spectra using H$_{2}$O band indices and aggregated by \citet{Hillenbrand2013} in their Table 3.  Some of these targets had been observed in multiple surveys with different molecular indices.  For these targets, we preferentially assign the optical spectral type over IR spectral type due to the larger uncertainty in IR data associated with variability in young stars as described in \citet{Simon_2020}.  For the 15 targets with a range of spectral types listed, we use the median spectral type, e.g. for M6-M7 we adopt M6.5.  This results in a total target list of 103 very low mass and brown dwarf ONC members.

From the list of ONC members with spectral types M6 - M9.5, we first limited our source list to those that appeared in the area covered by the HST Treasury Program of the ONC, resulting in 94 sources remaining.  Then, we limited our source list to those above a signal-to-noise (S/N) minimum (see Eq. 1) of 15 (the equivalent to a few hundred counts) as it is challenging to detect companions around faint primaries (see Sec. 2.4 and Fig. 1).  This minimum was set for each broadband filter meaning that as long as a target had a S/N of at least 15 in all its images in a single filter, it was included as a target in the sample (84 sources remaining).  Lastly, we limited our source list to objects with a maximum pixel value below 55,000 counts, which excludes sources affected by saturation, analogous to the procedure of AK06.  These filtering processes left the sample with 79 sources that were within the field of view and had sufficient S/N in at least one broadband filter to be analyzed with the double-PSF fitting tool.

For a source with the maximum pixel value equal to 55,000 counts, the entire PSF contains $\sim$ 250,000 counts.  This flux corresponds to a 0.2 \(\textup{M}_\odot\) and a 0.03 \(\textup{M}_\odot\) object in the F435W and F850LP filters respectively with no extinction, and a 0.07 \(\textup{M}_\odot\) object in the F850LP filter with a moderate amount of extinction (e.g. A$_{v}$ = 3).  The lower bound of our sample (S/N = 15, $\sim$ 500 total counts) corresponds to a 0.017 \(\textup{M}_\odot\) and a 0.005 \(\textup{M}_\odot\) object in the F435W and F850LP filters, respectively, with no extinction and a 0.05 \(\textup{M}_\odot\) object and a 0.007 \(\textup{M}_\odot\) object with a moderate amount of extinction. Therefore, higher mass and low extinction brown dwarfs are more likely found in the bluer filters while lower mass and higher extinction brown dwarfs are more likely found in redder filters, see Fig. \ref{fig:CMD}.

We then ran the double-PSF binary algorithm on our sample and set a chi squared threshold to indicate a reasonable fit to that in which a six-parameter model has a p-value equal to 0.1, $\chi^{2}_{\nu}$ = 1.774.  Below this value, we do not reject the null hypothesis that the data could have come from our model.  This resulted in a sample of 75 members found with S/N $\geq$ 15 and $\chi^{2}_{\nu}$ $\leq$ 1.774 in at least one broadband filter.  Sources can be excluded from the sample based on goodness of fit criteria due to either the presence of a higher order companion or due to high differential nebulosity.  In our sample, only four sources were excluded due to either being within a proplyd (one target) or in a region of high nebulosity (three targets).


\subsection{Completeness}\label{subsec:completeness}
\begin{figure*}[h]
\gridline{\fig{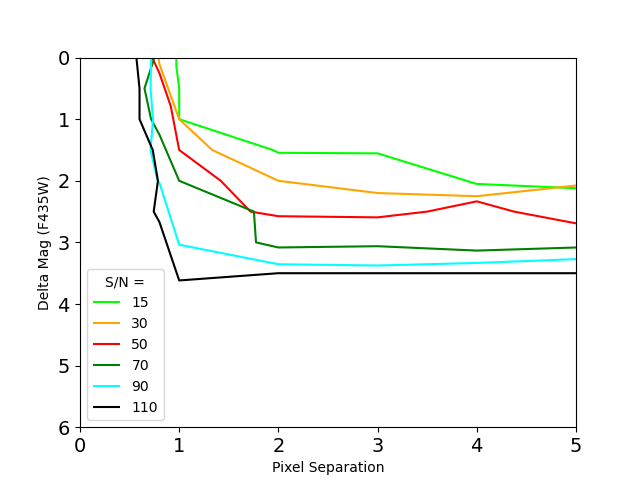}{0.49\textwidth}{a) F435W}
    \fig{completeness_lines_f555w_98_revised.png}{0.49\textwidth}{b) F555W}}

\gridline{\fig{completeness_lines_f775w_98_use.png}{0.49\textwidth}{c) F775W}
    \fig{completeness_lines_f850lp_98_revised.png}{0.49\textwidth}{ d) F850LP}}

\caption{99\% recovery lines for all S/N tested in the four broadband filters of ACS used in the HST Treasury Program of the ONC.  The different colors represent the recovery ability of our code for various S/N values.  As expected, our code can recover fainter and closer companions around brighter targets.  The angular resolution of our algorithm also improves as the S/N of the target increases.  Above this line, we are confident that our double-PSF fitting algorithm is recovering nearly all binaries and our method is complete.}
\label{fig:completeness}
\end{figure*}

First, we will characterize the ability of our binary PSF-fitting method to recover known binary systems before explaining how we determine whether or not a given double-PSF fit is representative of a true binary detection.  The sensitivity of this method is a function of the S/N of the target, the separation between primary and companion, the difference in magnitude between the two, and the filter in which the data were taken.  The S/N of a target is dependent on both the mass of the target and the interstellar extinction to the target. For two objects of the same mass,  the object with higher extinction will have a correspondingly lower S/N, and it will be more difficult to recover faint companions around that source. Conversely, it will be easier to recover faint companions around the source with less extinction because the source is brighter and therefore there is a larger contrast where we can see fainter objects between the source and the background limit, see Fig. \ref{fig:completeness}.

In Paper I, we determined our ability to recover known binary systems by fitting our double-PSF model to many artificially created binaries from the AK06 PSF libraries (see Sec. 2.2) of specific separations, specific differences in magnitude, and random position angles relative to the primary, including Poisson noise to simulate a generic background scene.  If the binary best fit separation parameter was within 0.2 pixels of the known separation and the best fit $\Delta$mag was within 0.2 mags of the known $\Delta$mag, that source would be defined as recovered.  We then set the detection limit as the location in $\Delta$mag and separation space where we recovered 90\% of the companions.  However, this cutoff of 0.2 mags in $\Delta$mag and 0.2 pixels in separation is arbitrary.

In this paper, we take a different approach to determine the binary recovery parameter space, described below.  We start by making artificial binaries from the AK06 PSF models as done in Paper I.  As before, the $\Delta$mag sampled were dependent on the S/N.  For higher S/N artificial binaries, we sampled integer (1.0-5.0) pixel separation binaries with $\Delta$mag down to 6.0 mags and sub-pixel (0.4, 0.6, and 0.8) separation binaries with $\Delta$mag down to 4.0 mags.  For lower S/N artificial binaries, we sampled in a similar fashion, but down to lower contrast (e.g. 4.5 mags for integer pixel separation binaries of S/N = 30).  This is again due to the fact that the range in $\Delta$mag space in which we can detect companions is directly related to the S/N of the target.  The S/N  is set by the mass and age of the target and the interstellar extinction to the target.  Within each bin of separation and $\Delta$mag, we created 500 binaries and ran the double-PSF code on these sources.


The S/N ratio (eq. \ref{sn}) of each source was calculated over all pixels in the 21x21 pixel stamp that encompasses each source with the mean of the local background subtracted as described in Sec. 2.2.  The total error per pixel was the summed in quadrature errors of the source photon noise, the standard deviation in the background, the error in the mean of the background, the dark current, and the read noise.

\begin{equation}
    S/N = \frac{\sum\limits_{i=1}^{N}  (data_{i} - \overline{bkgd})}{\sqrt{\sum\limits_{i=1}^{N}\sigma^{2}_{i}}}
\label{sn}
\end{equation}

In our double-PSF fitting code, the AMOEBA routine samples the six binary parameter space as described in Sec. \ref{subsec:model}.  We save each step taken by AMOEBA and its corresponding $\chi^{2}_{\nu}$ value.  Because we created these artificial binaries, we know the true binary parameters values for each binary.  Therefore, we can evaluate the goodness of fit of the double-PSF model at the exact value of the known binary parameters.  If that $\chi^{2}_{\nu}$ value is $>$ some critical value, then we can consider that binary not recovered by our method.  We select a critical value for the $\chi^{2}_{\nu}$ distribution corresponding to a certain probability that the known binary parameter values come from the model that was fit.  Here, we select the critical value ($\chi^{2}_{\nu}$ = 1.21) where there is a 99\% probability that a $\chi^{2}_{\nu}$ value less than this critical value for a given set of parameters could be drawn from distribution of our model fit parameters.  If the double-PSF fitting routine is complete for binaries of a given separation and $\Delta$ mag, we expect to recover 99\% of the 500 binaries within that bin, showing that our method can accurately recover these binaries.  We evaluate this metric for every artificial binary created within each bin in separation and $\Delta$mag space for all S/N values of all filters.  When the method begins to become incomplete, this recovered fraction will drop below 99\%.  This cutoff value of 99\% ensures that we expect to have failed to recover $<$ 1 binary within this study of 75 sources.  Fig. \ref{fig:completeness} shows the 99\% recovery line for each S/N tested for each filter used.  This curve serves to define where we are confident our sample is complete.  

Importantly, the completeness of this method varies for different S/N values.  A companion separated from the primary by three pixels with a $\Delta$mag of three mags will be detectable for all targets with S/N $\geq$ 90 in all filters, but will be undetectable around targets with S/N $\leq$ 70 in all filters.  Therefore, the detectable mass ratios will depend on the S/N of each target (i.e. its mass and interstellar extinction), discussed more in depth in Sec. \ref{subsec:compfrequency}.





\begin{figure*}[ht!]
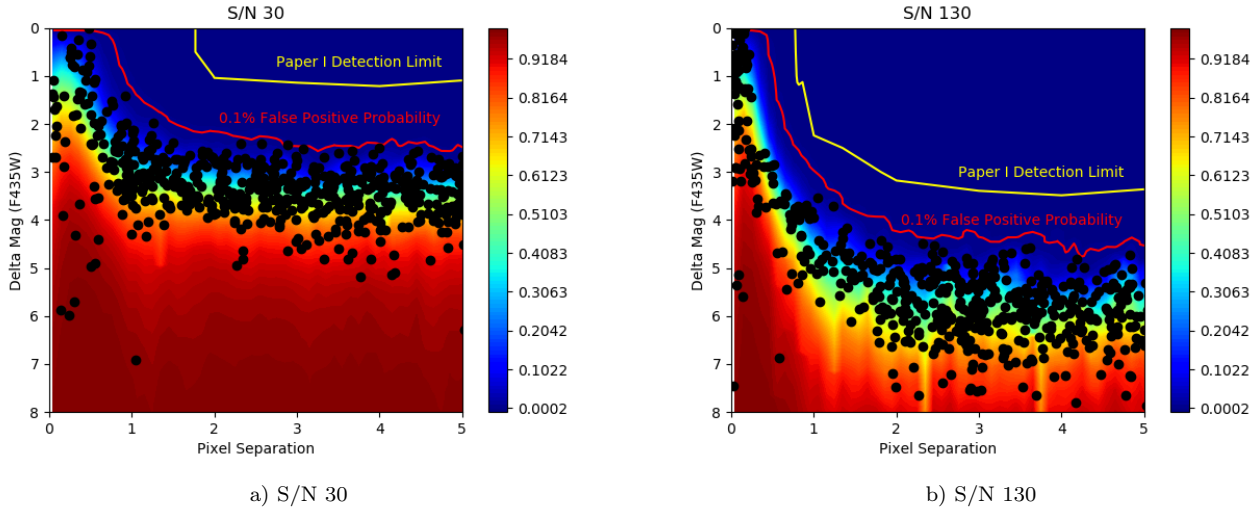

\gridline{\fig{Figure_2_FalsePositivesF435W_SNR30_forpaper.png}{0.49\textwidth}{a) S/N 30}
    \fig{Figure_2_FalsePositivesF435W_SNR130_forpaper.png}{0.49\textwidth}{b) S/N 130}}
\caption{False positive probability maps in separation and $\Delta$mag space.  The color scale represents the false positive probability where larger differences in magnitude correspond to a higher likelihood of a false positive fit.  We show two examples in the F435W filter with low S/N (30) and high S/N (130).  Over-plotted are the 0.1\% false positive probability lines and the 90\% completeness lines from Paper I.  The black circles represent the best fit binary parameters from the double PSF-fitting algorithm to the 1000 artificial single stars.  All 1000 best fits fall below the 0.1\% positive probability line, indicating that fits above this line are inconsistent with a single source.}
\label{fig:FP_HeatMaps}
\end{figure*}

\subsection{False Positive Analysis} \label{subsec:falsepositives}
In addition to characterizing our completeness, we must also investigate the reliability of our fitting code.  We can utilize false positive information to define the region of high probability detections.  A false positive occurs when our binary fitting code converges on a binary fit to a single star, but with a $\chi^{2}_{\nu}$ $\leq$ 1.774 (as detailed in Sec. \ref{subsec:subsample}).  In Paper I, false positives occurred significantly below the previously defined detection limit, and therefore nothing claimed as a detection could be confused for a false positive.  In this paper, we use the separation and $\Delta$mag values from the binary fits to artificial singles in order to determine the region in which our algorithm converges on a best-fit binary model, but the source is not a true binary.

The AMOEBA routine samples the six dimensional binary parameter space, fitting various binary models each with a calculated $\chi^{2}_{\nu}$, until converging.  When fitting to a single, our double-PSF fitting code will fit the primary PSF accurately while the secondary PSF is fit to the residuals of the primary PSF or to the background pixels.  Similar to the recovery of artificial binaries, the location in separation vs. $\Delta$mag space of false positive fits depends on the S/N of the source and the filter used, e.g. fitting to background pixels occurs at lower contrast around lower S/N targets.  

We created 1000 artificial single stars built from the models in the AK06 PSF libraries for each filter and specified S/N, adding Poisson noise to simulate a generic background scene.  Then, we ran our double PSF-fitting routine on each of the artificial single stars, using the best fit from a coarse grid search in the six dimensional binary parameter space as an input to the downhill simplex algorithm (same process as for fitting binaries).  

We use all steps and their associated $\chi^{2}_{\nu}$ within each fit to all 1000 single stars of a given S/N and filter in order to characterize the distribution of expected fits to single stars of the given S/N and filter.  First, we calculated the chi squared probability, P($\chi^{2}$), for each fit AMOEBA makes to each artificial single while stepping to reduce the $\chi^{2}$: 

\begin{equation}
    P(\chi^{2}) \propto \chi^{2^{(dof/2)-1}} * e^{-\chi^{2}/2}        
\label{chi2prob_dist}
\end{equation}

where \emph{dof} is the degrees of freedom equivalent to the number of data points (i.e. pixels) minus the number of free parameters in the model (the six binary parameters).  From the P($\chi^{2}$) equation, we calculate the P($\chi^{2}$) of each step while fitting to each artificial single star.

Then, we define the region of interest where the code is likely to fit the binary model to the data (whether it is fitting to a real companion, the background, or the residuals of the primary) in terms of separation, centered on the primary, from 0 - 10 pixels and $\Delta$mag from 0 - 10 mags with bin sizes of 0.1 pixels and 0.1 mags.   Within each bin, we select the best fit to each single, i.e. highest P($\chi^{2}$), and normalize the resulting distribution to one.  This creates a probability distribution in separation and $\Delta$mag space for the double-PSF fitting routine to an artificial single star.  We repeat this process for all 1000 artificial single stars for a given S/N and filter, which results in 1000 separate probability distributions of binary fits to artificial single stars.  We sum all of these 1000 distributions, normalizing the result to one, which results in one distribution of all of the binary fits to single stars with the same S/N and in the same filter. This process samples the entire region of interest and allows us to determine where in separation and $\Delta$mag space the code is likely to converge if it is in fact fitting to a single star.  Lastly, we sum vertically in $\Delta$mag space for each separation bin to find the value below which a certain percentage of the false positive fits will lie.  This process is performed for all filters and all S/N values.

In Fig. $\ref{fig:FP_HeatMaps}$, we show two examples of this in the F435W filter for S/N = 30 and 130.  The color scale corresponds to the false positive probability of any given separation bin below which the corresponding fraction of the false positive fits will lie.  We plot the 0.1\% false positive probability in both figures to show the value below which 99.9\% of false positive fits lie.  The binary best fit to a single star will only converge 0.1\% of the time above this value.  Black circles represent the best fit binary parameters to the 1000 artificial singles, all of which lie below the 0.1\% false positive probability line.  The mean $\chi^{2}_{\nu}$ of the 1000 best fits to these single stars of a given S/N and filter is 0.99, indicative of no unmodeled noise terms.  Binary fits above this value are likely binary detections.  We define this as the reliability of our method, where fits with a false positive probability $>$ 0.1\% are unreliable.  The corresponding curves are shown in Fig. \ref{fig:FPRateCurves} for each filter and S/N.

Fig. \ref{fig:SN90_ex} gives an example of the 0.1\% false positive probability curve for a S/N=90 target in the F435W with the completeness line for binaries of the same S/N in the F435W filter overplotted.  It is evident that the completeness limit and the 0.1\% false positive probability curve are overlapping.  This is the case across all S/N values and filters.  These results lead us to conclude that our sample is complete down to the 0.1\% false positive probability line, and we do not expect to miss any real binaries above this limit.  Therefore, we do not have to correct for incompleteness over the sensitivity of our survey.

Additionally, we compared the results of binary fits to artificial singles to binary fits to real singles within one filter and one S/N as an example of our method.  We identified sources within the data that had a S/N $\sim$ 30 in the F435W filter that were well fit with a single PSF.  Each source was run through the binary-fitting code, and we repeated the process above to derive the false positive maps similar to Fig. \ref{fig:FP_HeatMaps}.  In Fig. \ref{fig:TestonRealSources}, we show the 0.1\% false positive probability line when fitting our binary PSF model to real singles and artificial singles.  Their overlap, particularly at small separations, indicates that the empirical PSFs of AK06 accurately model the data.

\begin{figure*}[h]
\gridline{\fig{Figure_3_F435W_falsepositives_allsnr.png}{0.49\textwidth}{a) F435W}
    \fig{Figure_3_F555W_falsepositives_allsnr.png}{0.49\textwidth}{b) F555W}}

\gridline{\fig{Figure_3_F775W_falsepositives_allsnr.png}{0.49\textwidth}{c) F775W}
    \fig{Figure_3_F850LP_falsepositives_allsnr.png}{0.49\textwidth}{ d) F850LP}}

\caption{The 0.1\% false positive probability curves for all specified S/N values and the four broadband filters as described in Sec. \ref{subsec:falsepositives}.  The circles represent the detected companions listed in Tables \ref{table1} and \ref{table2} and are color-coded based on S/N to match the corresponding reliability limit of the source.}
\label{fig:FPRateCurves}
\end{figure*}

\begin{figure}[ht!]
\gridline{\fig{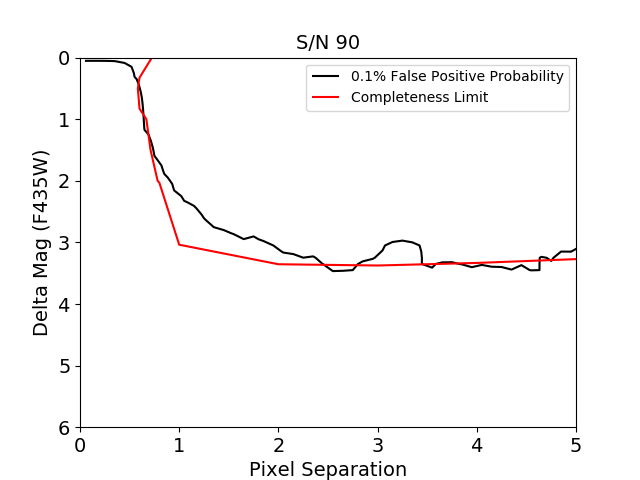}{0.49\textwidth}{}}
\caption{Plotted are the completeness limit and 0.1\% false positive probability curve associated with targets of S/N=90 in the F435W filter.}
\label{fig:SN90_ex}
\end{figure}

\begin{figure}[ht!]
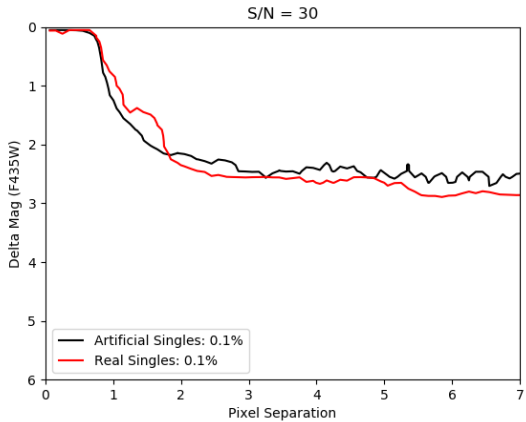

\gridline{\fig{Figure_5_SNR30_comparison.png}{0.45\textwidth}{}}
\caption{0.1\% false positive probability lines for actual single stars and artificial singles built from the empirical PSFs of AK06 for S/N = 30 targets.}
\label{fig:TestonRealSources}
\end{figure}

From the false positive maps, we can then calculate a false positive probability for each target based on its resulting chi squared probability distribution in separation and $\Delta$mag space from the binary fitting routine.  As above, we create a probability distribution in separation and $\Delta$mag space from all the steps taken within the double-PSF fitting routine to a given source.  We multiply this probability distribution by the corresponding false positive distribution of the same S/N (as exemplified in the heat map in Fig. \ref{fig:FP_HeatMaps}) and take the sum to derive the false positive probability of the fit to the data (here one image of one source).  Most of our targets appear in multiple images of the Treasury Program data as well as in multiple filters.  To calculate the false positive probability of a source with multiple images within a filter, we combine the false positive probabilities within each image of the filter:

\begin{equation}
    FP_{filter1} = FP_{i}*FP_{i+1}*...*FP_{N}
\label{fp_prob_onefilter}
\end{equation}

where FP$_{filter1}$ is the false positive probability of one target within filter1 and FP$_{i}$ is the false positive probability of one image of that target within the same filter.  We then use the false positive probability derived from all images within the filter, FP$_{filter1}$, to determine whether or not that target has a detected companion or if it is classified as a single.  Here, we define sources as detected binaries if the false positive probability is $\leq$ 0.1\%.

If the target appears in multiple images within multiple filters, we can also combine those false positive probabilities for a global false positive probability across all filters:

\begin{equation}
    FP_{total} = FP_{filter1}*FP_{filter2}*...*FP_{filterN}
\label{fp_prob_multifilter}
\end{equation}

where FP$_{total}$ is then used to determine whether or not that target has a detected companion.

This combination of false positive probabilities requires the code to be fitting to the same feature within all images and in all filters.  Therefore, we must confirm that we are identifying the same feature in all the binary fits to every image of the same source.  We can verify this by measuring the similarity between the chi squared probability distributions that result from the binary fits to a given target in multiple images.  We measure this similarity through the Bhattacharyya coefficient (BC).  This statistic quantifies the amount of overlap between any two distributions, and is defined as:

\begin{equation}
    BC = \sum\limits_{i=1}^{N} \sqrt{p_{i}*q_{i}}
\label{bc_coeff}
\end{equation}

where there are N bins in our multivariate parameter space and p and q are the two probability distributions in question for the two different images.

This metric is bound between 0 and 1 where 0 indicates no overlap, and 1 indicates complete overlap.  We tested the extent to which this metric would distinguish between two distributions by creating two identical normal distributions and varying their centers, moving them further apart.  Then, we calculated both the Bhattacharyya coefficient and the sigma difference between the two distributions.  We found that at a 3$\sigma$ difference, the Bhattacharyya coefficient equals 0.1.  Therefore, there is 99.7\% confidence that a Bhattacharyya coefficient below 0.1 is indicative of two distinct distributions.  We use this critical value to confirm the similarity between probability distributions in two images (if BC $>$ 0.1) in order to combine their false positive probabilities for a global false positive probability.  When a Bhattacharyya coefficient $<$ 0.1 occurs, we select the image with the lowest $\chi^{2}_{\nu}$ to calculate the false positive probability and ignore the other image.

This is done among all images of the same source in the same filter where we calculate the BC between images in the three dimensions associated with the companion: separation, position angle, and difference in magnitude.  The false positive probability of that target within a filter (FP$_{filter1}$) will then be used to assess whether or not there was a detection, where we define a false positive probability $>$ 0.1\% as a non-detection.  If that source appears in multiple filters we calculate the BC in the two dimensions that are consistent across all filters: separation and position angle.  The false positive probability FP$_{total}$ is then used to assess binarity in the same way that FP$_{filter1}$ is used to assess binarity if the target appears in one filter.

\section{Results} \label{sec:results}
\subsection{Detections} \label{subsec:detections}
Our sample consists of 75 ONC members that meet the selection criteria defined in Sec. \ref{subsec:subsample} and are all classified as M6-M9.5.  After analyzing each source in all images and filters, we detected 8 candidate binaries with false positive probabilities $<$ 0.1\% and binary fits with $\chi^{2}_{\nu}$ $< $ 1.774.  Two of these candidate companions were only identified in one filter (one in F555W and one in F775W).  Six of these candidate companions were identified in at least two filters in all images in which they appeared.  All detections were found in at least two images.

\begin{figure*}[]
\gridline{\fig{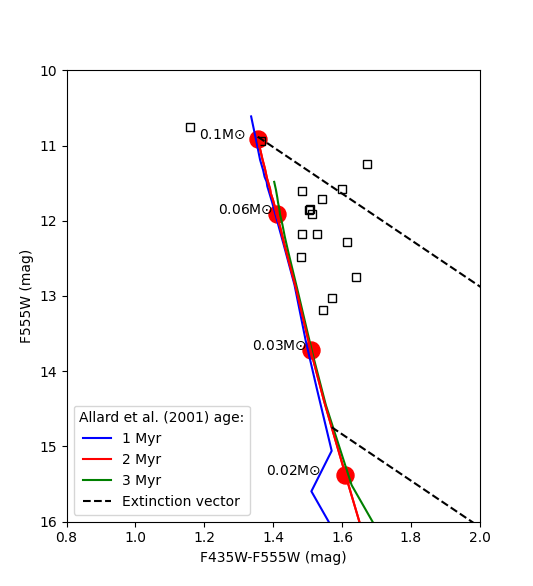}{0.5\textwidth}{Color-magnitude diagram in F435W and F555W}
    \fig{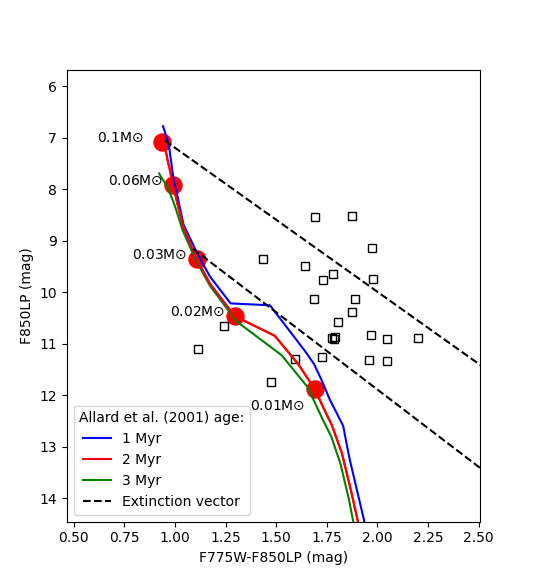}{0.5\textwidth}{Color-magnitude diagram in F775W and F850LP}}
\caption{Color-magnitude diagrams for the targets (open black squares) in our sample that have S/N $\geq$ 30 in both filters represented in either diagram, corrected for distance (400 pc).  The solid lines denote the 1, 2, and 3 Myr isochrones from \citet{Allard2001}, the dashed black lines denote the effect of interstellar extinction.  The effective wavelength of each filter is used to calculate extinction with the procedure of \citet{Cardelli1989}  Filled red circles are included to note parts of the isochrone corresponding to specific masses, 0.02 - 0.1 M$_{\odot}$ on the left, 0.01 - 0.1 M$_{\odot}$ on the right.}
\label{fig:CMD}
\end{figure*}

In general, there are three potential outcomes from the analysis in Sec. \ref{subsec:falsepositives}: 1) a well-detected binary with a false positive probability $<<$ 0.1\%; 2) a binary fit with a false positive probability $>>$ 0.1\%; and 3) a fit with a false positive probability near 0.1\% and binary parameters poorly constrained due to the influence of the primary PSF residuals or the background.  Of the eight detections, seven were well-detected binaries in at least one filter in which they appeared, well above the completeness limit and reliability threshold.  Only one detection (J05350957-0519426 at 0.028") was observed near the 0.1\% false positive probability for a single image in a single filter with poorly constrained binary parameters.  In all cases where the companion was bright enough to detect above the reliability threshold and the images were unsaturated, we detected the companion in all images of all filters.  For the companion only identified in the F555W filter (J05351624-0528337), the primary was too faint to observe in the F435W filter, and the images were saturated in F775W and F850LP.  For the companion only identified in the F775W filter (J05350957-0519426), the primary was too faint to detect in F435W and F555W, and the companion was too close in separation to detect in the F850LP filter (assuming the $\Delta$ mag in F850LP corresponded to the estimated mass ratio of the binary).

The projected separations of the candidate companions range from 11 - 121 AU (0.028" - 0.30", or 0.55 - 6.06 pixels on ACS).  Two of these candidate companions, both with separations $>$ 100 AU, were previously detected by \citet{Reipurth2007}, and one of these same two was identified in Paper I where the primary mass estimate has been corrected based on the \citet{Hillenbrand2013} spectral type.  One other candidate companion (J05351624-0528337) was previously identified in Paper I.  Because this target only appeared in one filter, we could not estimate our own extinction value so we used the extinction estimate from the most recent spectroscopic observation.  We adopt A$_{v}$ = 4.5 from \citet{Kounkel2018}, which, combined with the spectral type of \citet{Hillenbrand2013}, resulted in a primary mass of 0.15 M$_{\odot}$, larger than the upper limit of our target sample.  Therefore, this target was excluded from our sample, leaving us with 7 candidate very low mass binaries.  The remaining five detections are new.  

In Table \ref{table1}, we present the observed color information derived for each source, i.e. the combined magnitude of the system and difference in magnitude between the primary and secondary in each filter.  All binary detections have a $\Delta$ mag $<$ 2 mag across all filters, shown in Table \ref{table1}  and Fig. \ref{fig:FPRateCurves}.  Our method is sensitive to companions at 4 mags in contrast, although this is only the case for high S/N targets.  For lower S/N targets, we are sensitive to lower contrast companions.  See Sec. \ref{subsec:compfrequency} for a discussion on the sensitivity of a large sub-set of the sample and Sec. \ref{subsec:SampleResult} for our statistical approach in determining that we do not have a bias against low mass ratio binaries.  We also display the photometric information in two color-magnitude diagrams (CMDs) shown in Fig. \ref{fig:CMD}.  We limit the sources included to those with S/N $\geq$ 30 in both filters that the CMDs cover.  Included in the plots are the 2 Myr isochrone of \citet{Allard2001} and extinction vectors.  These plots show that higher mass and/or lower extinction objects are more often found in the bluer filters while lower mass and/or higher extinction objects are more often found in the redder filters.  In Table \ref{table2}, we present the physical parameters of each binary (masses, separation, and position angle) and the measured values from our analysis (e.g. S/N and $\chi^{2}_{\nu}$).  

We estimate the extinction of each source with multi-filter data using the observed flux of the primary with ACS, the intrinsic model flux of the primary derived from \citet{Allard2001} based on the spectral type of the target, and the extinction law of \citet{Cardelli1989}.  For one target (J05350739-0525481), the estimated extinction is -1.4.  This target was observed to have Ca II triplet emission lines in \citet{Hillenbrand2013} which are indicative of accretion.  Shocks from accretion onto an object along magnetic field lines will cause excess emission in blue/UV wavelengths \citep{Azevedo2006}.  This could cause a source to have higher flux in bluer filters, and could account for the negative extinction estimate.  Since this negative extinction is non-physical, we assume an extinction of zero similar to the treatment of \citet{Herczeg2014} and \citet{Kounkel2018}.

We calculate masses using the isochrones of \citet{Allard2001}, assuming that the derived extinction is the same for both the primary and secondary components of the binary.  The ONC has an observed age spread of $\sim$ 1-3 Myr \citep{Reggiani2011, Beccari2017}.  If we assume ages of 1, 2, and 3 Myr, our mass estimates vary by $<$ 20\% while the mass ratio estimates vary by $<$ 0.1.  Because we do not have individual age estimates of our detected binaries, we derived masses using the 2 Myr isochrone, the mean age estimated by \citet{Reggiani2011}. If instead we use the COND models of \citet{Allard2001}, all mass estimates change by $<$ 25\% while the mass ratio estimates vary by $<$ 0.15.

The physical separation is calculated assuming a distance of 400 pc to each target \citep{oriondistance} and applying a correction factor of 1.16 to convert from projected to physical separation in a statistical sense \citep{Dupuy2011}.  Fig. \ref{fig:Binaries_1} shows one image of each detection in the F435W and F555W filters in order of separation.  Fig. \ref{fig:Binaries_2} shows the same but for the F775W and F850LP filters.

\begin{figure*}[h]
\gridline{\fig{F435W_binarywcolorbar.png}{0.89\textwidth}{F435W Detections}}
\gridline{\fig{F555W_binarywcolorbaredit2.png}{0.92\textwidth}{F555W Detections}}
\caption{We show the binary detections of our survey in the F435W and F555W filters.  For each panel, the top image is the postage stamp of the \emph{HST} data, the middle image is our binary PSF best-fit model, and the bottom image is the residual.  The binaries are listed from left to right in order of increasing projected separation.  Each postage stamp is 21x21 pixels across, an angular size of 1.05"x1.05".  The images are displayed with an inverse hyperbolic sine stretch and the units are in counts. }
\label{fig:Binaries_1}
\end{figure*}


\begin{figure*}[]
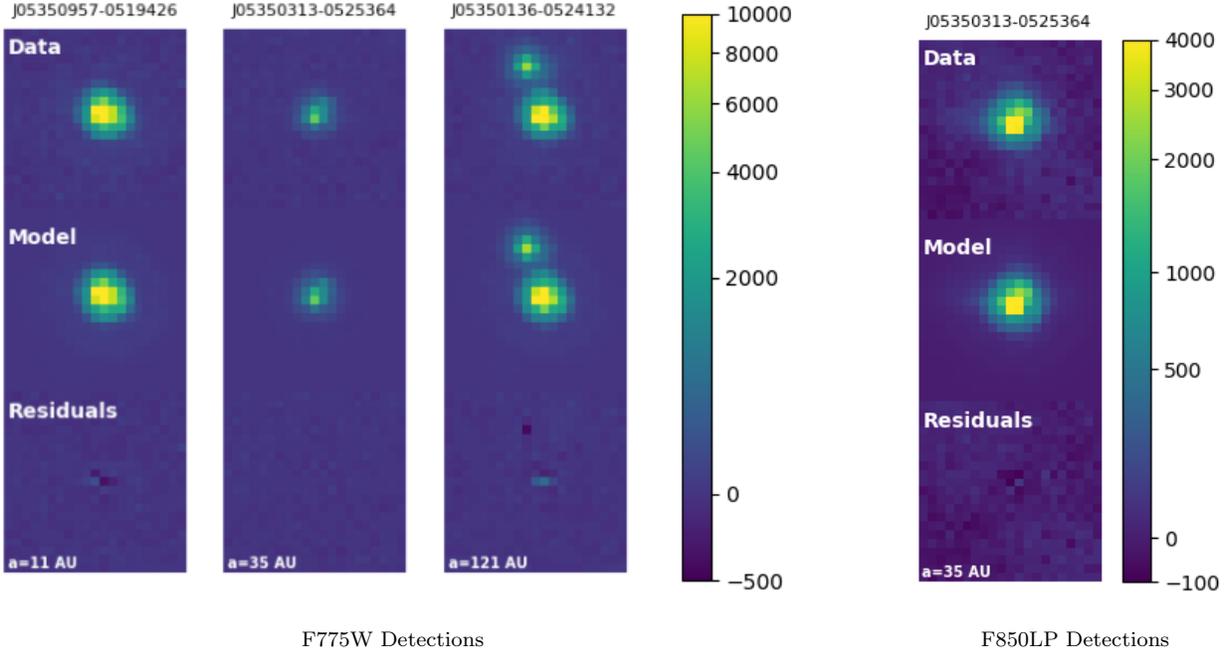

\gridline{\fig{F775W_binarywcolorbar.png}{0.6\textwidth}{F775W Detections}
    \fig{F850LP_binarywcolorbar.png}{0.24\textwidth}{F850LP Detections}}
\caption{Same as Fig. \ref{fig:Binaries_1} but for the F775W and F850LP filters.}
\label{fig:Binaries_2}
\end{figure*}

\begin{deluxetable*}{ccccccccc}
\tablenum{1}
\tablecaption{We list the observed magnitude information of each binary system.  In each set of two columns, we present the total magnitude of each system in each filter followed by the difference in magnitude between the primary and companion.  These values are the weighted mean calculated from all exposures of the target where the error is the 68\% confidence interval.  We use these values to estimate masses in Table \ref{table2}.  Where values are not listed, they did not fulfill the requirements listed in Sec. \ref{subsec:subsample}. \label{table1}}

\tablewidth{0pt}
\tablehead{
\colhead{Source \#} & \colhead{Total Mag} & \colhead{$\Delta$ mag} & \colhead{Total Mag} & \colhead{$\Delta$ mag} & \colhead{Total Mag} & \colhead{$\Delta$ mag} & \colhead{Total Mag} & \colhead{$\Delta$ mag} \\
\colhead{} & \colhead{(F435W)} & \colhead{(F435W)}& \colhead{(F555W)} & \colhead{(F555W)}& \colhead{(F775W)} & \colhead{(F775W)}& \colhead{(F850LP)} & \colhead{(F850LP)}}
\decimalcolnumbers
\startdata
  1 & 23.07 $\pm$ 0.04 & 0.09 $\pm$ 0.06 &  22.10 $\pm$ 0.03 & 0.52 $\pm$ 0.05 &  18.74 $\pm$ 0.01 & 1.48 $\pm$ 0.02 &  - &  -\\
  2 & 23.10 $\pm$ 0.04 & 0.71 $\pm$ 0.07 &  21.20 $\pm$ 0.02 & 0.62 $\pm$ 0.03 &  - & - &  - &  -\\
  3 & 19.92 $\pm$ 0.02 & 0.90 $\pm$ 0.04 &  18.76 $\pm$ 0.01 & 0.39 $\pm$ 0.04 &  - & - &  - &  -\\
  4 & 23.55 $\pm$ 0.03 & 1.31 $\pm$ 0.09 &  21.35 $\pm$ 0.01 & 1.33 $\pm$ 0.02 &  - & - &  - &  -\\
  5 & 20.92 $\pm$ 0.02 & 1.12 $\pm$ 0.25 &  19.24 $\pm$ 0.01 & 0.94 $\pm$ 0.08 &  - & - &  - &  -\\
  6 & - & - &  - & - & 20.67 $\pm$ 0.02 & 1.68 $\pm$ 0.05 & 18.89 $\pm$ 0.02 &  1.57 $\pm$ 0.05\\
  7 & - & - &  - & - &  18.80 $\pm$ 0.01 & 1.37 $\pm$ 0.57 &  - &  -\\
  8$^{a}$ & - & - &  22.3 $\pm$ 0.03 & 0.72 $\pm$ 0.06 &  - & - &  - &  - \\
\enddata
\tablenotetext{a}{This target is excluded from the final sample, see Sec. \ref{subsec:detections}.}
\end{deluxetable*}

\begin{deluxetable*}{ccccccccccc}
\tablenum{2}

\tablecaption{Candidate binaries with masses (\(\textup{M}_\odot\)), mass ratios (q), projected separations in arcseconds, physical separation in AU, and position angles in degrees.  To estimate masses, we used the 2 Myr isochrone of \citet{Allard2001}.  We assumed a distance of 400 pc \citep{oriondistance} and applied a correction factor of 1.16 to convert from projected separation to physical separation \citep{Dupuy2011}.  Binary parameters  are the weighted mean calculated from all exposures of the target where the error is the 68\% confidence interval.  Errors in the mass estimates come solely from errors in the photometry, and do not take into account the age spread in the ONC or evolutionary model used. Sources \#1 and 4 were previously identified in \citet{Reipurth2007}.  We also show the S/N of the target for the filter within which it is brightest along with the $\chi^{2}_{\nu}$ associated with that filter.  We do not show the S/N and $\chi^{2}_{\nu}$ within each filter, but they fulfill the requirements of Sec. \ref{subsec:subsample}.  \label{table2}}

\tablewidth{0pt}
\tablehead{
\colhead{Source \#} & \colhead{2MASS ID} & \colhead{M$_{prim}$} & \colhead{M$_{sec}$} & \colhead{q} & \colhead{A$_{v}$} & \colhead{Projected} & \colhead{Physical} & \colhead{PA (deg)} & \colhead{S/N} & \colhead{$\chi^{2}_{\nu}$} \\
\colhead{} & \colhead{} & \colhead{(M$_{\odot}$)} & \colhead{(M$_{\odot}$)} & \colhead{} & \colhead{(mag)} & \colhead{Sep. (arcseconds) } & \colhead{Sep. (AU)} & \colhead{(E of N)} & \colhead{} & \colhead{} }
\decimalcolnumbers
\startdata
  1 & J05350136-0524132 &  0.044 $\pm$ 0.002 &  0.036 $\pm$ 0.002 &  0.82 & 1.9 &  0.318 $\pm$ 0.009 & 148 $\pm$ 4 &  114 $\pm$ 0.5 &  335  & 1.12\\
  2 & J05345805-0529405 &  0.046 $\pm$ 0.002 &  0.037 $\pm$ 0.002 &  0.80 & 1.1 & 0.113 $\pm$ 0.006 & 52 $\pm$ 2 & 64 $\pm$ 0.8 &  167  & 0.94\\
  3 & J05350739-0525481 &  0.079 $\pm$ 0.003 &  0.065 $\pm$ 0.002 &  0.82 & 0.0$^{a}$ & 0.070 $\pm$ 0.003 & 32 $\pm$ 1 & 153 $\pm$ 0.8 &  307  & 0.75\\
  4 & J05351676-0517167 &  0.093  $\pm$ 0.006 &  0.050 $\pm$ 0.003 &  0.54 & 2.6 &  0.270 $\pm$ 0.006 & 124 $\pm$ 2 & 265 $\pm$ 1.2 &  87  & 0.94\\
  5 & J05351404-0525499 &  0.10 $\pm$ 0.015 &  0.063 $\pm$ 0.007 &  0.63 & 0.8 & 0.052 $\pm$ 0.009 & 24 $\pm$ 4 & 280 $\pm$ 1.6 &  197  & 0.91\\
  6 & J05350313-0525364 &  0.023 $\pm$ 0.001 &  0.012 $\pm$ 0.001 &  0.52 & 1.1 & 0.086 $\pm$ 0.005 & 40 $\pm$ 2 & 58 $\pm$ 1.0 &  141  & 1.01\\
  7 & J05350957-0519426 &  0.033 $\pm$ 0.005 &  0.021 $\pm$ 0.004 &  0.63 & 1.2$^{b}$ & 0.027 $\pm$ 0.007 & 13 $\pm$ 3 & 129 $\pm$ 2.8 &  263  & 0.98\\
  8$^{d}$ & J05351624-0528337 &  0.15 $\pm$ 0.010 &  - &  - & 4.5$^{c}$ & 0.17 $\pm$ 0.006 & 67 $\pm$ 2 & 183 $\pm$ 0.9 &  32  & 0.86\\
\enddata
\tablenotetext{a}{Our extinction estimate using photometry from the F435W and F555W filters was -1.4.  As explained in Sec. \ref{subsec:detections}, this estimate is non-physical and instead we assume an extinction of 0.0 following the procedure of \citet{Herczeg2014} and \citet{Kounkel2018}.}
\tablenotetext{b}{Extinction estimate taken from the source paper of the spectral classification \citep{Riddick2007} due to the identification of the companion in only one filter.  This value was used in order to derive mass estimates.}

\tablenotetext{c}{Extinction estimate taken from most recent spectral observation \citep{Kounkel2018} due to the identification of the companion in only one filter.  This value was used in order to derive mass estimates.}

\tablenotetext{d}{This target is excluded from the final sample used to calculate the binary frequency of the ONC as its estimated mass of 0.15 M$_{\odot}$ is above the upper bound to be included.}
\end{deluxetable*}

\subsection{Detection Limits} \label{subsec:detectionlimits}
With the improved data analysis method described in Sec. 2, we increased the sensitivity of our double-PSF fitting routine, achieving 1.5 times higher spatial resolution and detecting companions 1 magnitude fainter than reported in Paper I.  As stated in Sec. 2.5, we define the 0.1\% false positive rate as our reliability limit, consistent with our completeness limit.  For the highest S/N tested, 130, we can identify companions at separations down to 0.5 pixels (see Fig. 4) compared to 0.75 pixels using the previous detection limit of Paper I.  We can also find companions at 4.25 mags in contrast beyond 2 pixels, i.e. the background limit, compared to 3.25 mags with the detection limit of Paper I.  For reference, these enhancements allow the identification of companions at 10 AU compared to 15 AU for the brightest targets in the ONC.  For a 0.1 M$_{\odot}$ primary with an age of 2 Myr, we can identify companions with masses of 21 M$_{Jup}$ compared to 27 M$_{Jup}$ in our previous work in the F555W filter based on the \citet{Allard2001} evolutionary tracks.  For lower S/N targets, we have a similar enhancement in resolution and contrast as shown in Fig. \ref{fig:FP_HeatMaps}.


\subsection{Chance Alignments} \label{subsec:contaminants}
\begin{figure}[h]
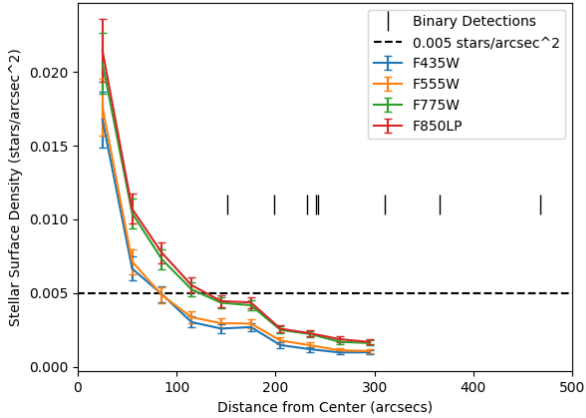

\gridline{\fig{Density_v_distance_all_wlines.png}{0.47\textwidth}{}}
\caption{Stellar density as a function of angular distance from Theta 1 Ori C.  We plot the stellar density for all four broadband filters used in this study, over-plot the 0.005 stars/arcsec$^2$ line, and over-plot the distance from each detection to Theta 1 Ori C.  Included are 1$\sigma$ error bars assuming the number of sources within each annulus is Poisson distributed.}
\label{fig:density}
\end{figure}

Before calculating the binary frequency in the ONC, we must first determine whether or not any of these detections are likely to be chance alignments.  There are three potential sources of contamination in binary surveys of star clusters: 1) foreground stars, 2) other cluster members, and 3) background sources.  We used the source list from \citet{Robberto2013} which contains all targets within the field of view (not just members) in each broadband filter to estimate the stellar surface density.  First, we define the center of the cluster at the location of Theta 1 Ori C, the most massive star in the cluster.  Then, we create 10 successive circular annuli each 30" further from the center than the last.  Within each annulus, we calculate the number of sources identified in \citet{Robberto2013} in the specified filter and calculate the stellar density.  We take the midpoint of the annulus as its distance from Theta 1 Ori C and plot the stellar density in each filter as a function of distance from the center of the ONC, as seen in Fig. \ref{fig:density}.

As expected, the stellar density is highest near the center, but flattens out around 0.001 stars/arcsec$^{2}$ after $\sim$ 250".  This result is consistent across all four broadband filters, with the redder filters having a higher density around 0.002 stars/arcsec$^{2}$.  In our survey, we search for companions within a 10 pixel radius from each target.  The detected binary closest to Theta 1 Ori C is 151" away.  The highest stellar density of all broadband filters in our survey at 150" from Theta 1 Ori C is 0.0045 stars/arcsec$^{2}$.  Using this stellar density as the upper limit to the expected density within the cluster, we expect 0.0035 chance alignments per target within this search radius and 0.26 chance alignments for all 75 targets in our sample. Although we searched a radius of 0.5" from the target, the widest detection was found at 0.3".  We expect to find 0.1 chance alignments to all 75 targets in the sample within a 0.3" radius.  Since we used the entire catalog of sources from \citet{Robberto2013} without membership information, this includes foreground and background contaminants.  Therefore, we conclude that all of the detections are likely physical binaries with no expected chance alignments.

\subsection{Binary Frequency} \label{subsec:compfrequency}
Before we can calculate a model independent binary frequency for 0.012 - 0.1 M$_{\odot}$ primaries in the ONC, we must define the common parameter space over which the detection method is sensitive to companions for a large sub-set of the sample, e.g. mass ratio and separation.  The initial sample is defined in Sec. \ref{subsec:subsample}, but the sensitivity to companions depends on the S/N of the targets and the filters in which those targets meet the requirements laid out in Sec. \ref{subsec:subsample}.  To calculate the achievable contrast within an image for a given source, we take the value from the 0.1\% false positive line shown in Fig. 4 at some specified separation.  Then, we calculate the mass of a companion corresponding to that highest achievable contrast from the \citet{Allard2001} evolutionary models, assuming the primary has the mass associated with the defined spectral types of \citet{Hillenbrand2013}, an age of 2 Myr, and the extinction of the companion is the same as the primary.  For any given source that appears in multiple images in one filter, we calculate the S/N of the target in each image and use the minimum S/N to determine the achievable contrast for that source within all images of that filter.  If the source appears in other filters, we repeat the S/N calculations as before and determine the achievable contrast within each filter.  Then, we select the lowest detectable companion mass achievable in a single filter as the companion detection threshold for a given target.  Lastly, we calculate the mass ratio, m$_{comp}$/m$_{prim}$, for each source to determine the common mass ratio range for most of the sample at a given separation.

As shown in Fig. \ref{fig:FPRateCurves}, the sensitivity of this method becomes background limited beyond about 2 pixels (0.1") in separation.  At the distance of the ONC ($\sim$ 400 pc), this corresponds to a projected separation of 40 AU.  Many past surveys of field brown dwarfs \citep[e.g.][]{Close2003, Gizis2003,Reid2006,Fontanive2018} show that the separation distribution can be modeled by a log-normal distribution with a peak somewhere between 3 - 10 AU, depending on primary mass and resolution limits.  Additionally, the companion mass ratio distribution (CMRD) of field binaries with brown dwarf primaries is strongly peaked at one \citep[e.g.][]{Reid2006}.  Because the separation distribution peaks at small separation and half of our detections are within 40 AU, we want to compare our detections to as much of the expected field population as possible and opt for a closer separation for common sensitivity.  We also opt for a moderate mass ratio limit to maximize the number of targets included in the sub-sample since the field surveys are not sensitive to companions at low mass ratios and to include as many detected companions as possible.  

Taking all of this into account, we define the range of common sensitivity for projected separations 20 - 200 AU and from a mass ratio of 0.5 - 1.0 which is applicable to 50 targets out of our initial sample of 75.  We are assured that the double-PSF fitting routine can detect all companions to these 50 targets (defined as the sub-sample) over mass ratios of 0.5-1.0 and projected separations of 20-200 AU.  Six of the seven detections in this survey had estimated mass ratios $>$ 0.5 and projected separations $>$ 20 AU as well as high enough S/N for the code to detect companions above a mass ratio of 0.5 at 20 AU.  One detection was at a projected separation of 11 AU and therefore excluded from this analysis.  Out of this analysis, we calculate a binary frequency of 12$_{-3.2\%}^{+6.0\%}$ for brown dwarf primaries (M$_{prim}$ $\leq$ 100 M$_{Jup}$) in the ONC over projected separations 20 - 200 AU and mass ratios 0.5 - 1.0.  Errors on the binary frequency are estimated from binomial statistics where we define the upper and lower bounds based on the 68\% confidence range of the binomial distribution \citep{Burg2003}.



\section{Discussion} \label{sec:discussion}
\subsection{Comparison to the Galactic Field} \label{subsec:SampleResult}
We next compare the observed binary frequency in the ONC to that of the Galactic field over the common sensitivity range of our sample, described in Sec. \ref{subsec:compfrequency} .  \citet{Reid2006} carried out an L-dwarf multiplicity survey in the Galactic field using the Near-Infrared Camera and Multi-Object Spectrometer (NICMOS).  Their sample consists of field objects with masses $\sim$ 0.04 - 0.09 M$_{\odot}$, where higher order multiples are very rare, making it an appropriate sample to compare to the binary results of our ONC survey, where M$_{prim}$ = 0.012 - 0.1 M$_{\odot}$.  We characterize the binary population of the Galactic field using functional forms of the observed binary parameters: the CMRD modeled as a power law (eq. \ref{qdistribution}) and the physical separation distribution  modeled as a log-normal (eq. \ref{adistribution}).  \citet{Reid2006} used a Bayesian analysis to estimate the values of the functional form variables that describe the observed field L-dwarf binary population: 1) binary frequency (C$_{n}$), 2) mean of the physical separation distribution ($\overline{a}$), 3) width of log-normal physical separation distribution ($\sigma$), and 4) power-law index ($\beta$) of the CRMD.  Their Fig. 7 shows the posterior probability distributions of this four-parameter model with best fits: C$_{n}$ = 24\%$_{-2}^{+6}$, $log(\overline{a})$ = 0.8$_{-0.12}^{+0.06}$, $\sigma$ = 0.28 $\pm 0.4$, and $\beta$ = 3.6 $\pm$ 1.  

\begin{equation}
    \frac{dN_{1}}{dq} \propto q^{\beta}
    \label{qdistribution}
\end{equation}

\begin{equation}
\frac{dN_{2}}{da} = \frac{1}{a \sqrt{2\sigma^{2}}} e^{-\frac{(log(a)-log(\overline{a}))^{2}}{2\sigma^{2}}}
    \label{adistribution}
\end{equation}

\begin{equation}
    CF = C_{n}*\int_{q_1}^{q_2} \frac{dN_{1}}{dq} \int_{a_1}^{a_2} \frac{dN_{2}}{da}
    \label{cf}
\end{equation}

We estimated the expected brown dwarf binary fraction of the Galactic field using eq. \ref{cf}.  First, we sampled each of the four posterior distributions of \citet{Reid2006} 10$^{6}$ times.  Then, we integrated eq. \ref{cf} over the common mass ratio and physical separation limits, 0.5 $<$ q $<$ 1.0 and 23 $<$ a $<$ 230 AU to determine the binary frequency of each sample.  We took the mean of this result as the expectation value for the binary frequency.  To approximate error bars, we defined the upper and lower bounds as the 68\% confidence interval of this resultant binary frequency distribution.  We derive the inner and outer bound of the separation by converting projected separation to physical separation \citep{Dupuy2011}.  These calculations result in an expected field binary frequency of 0.9\%$\pm 0.6$\%.  

In order to compare the binary frequency of the Galactic field to that of the ONC, we took the binomial distribution formalism of \citet{Burg2003}, described in Sec. \ref{subsec:compfrequency}, and integrated from 0 to the resulting binary frequency from each sampling above.  This statistic determines the posterior probability of the field binary frequency being the real frequency of binaries in the ONC population, i.e. the probability that Galactic field binary population comes from this ONC binary population.  Then, we took the mean probability of all the sampling as the expectation value that the field sample binary frequency derived by \citet{Reid2006} is consistent with our ONC results.  This procedure resulted in a mean probability of 10$^{-6}$.  This is evidence for an excess of VLMB in the ONC compared to the Galactic field over 0.5 $<$ q $<$ 1.0 and 23 $<$ a $<$ 230 AU.

\citet{Reid2006} calculate a strong preference for high mass ratio VLMBs in the Galactic field with a CMRD power-law of 3.6.  We sought to determine whether we could statistically differentiate the observed VLMB mass ratio distribution of the ONC from that of the model of the Galactic field.  To investigate this possibility, we sampled the \citet{Reid2006} posterior probability distribution of the power law index for the mass ratio distribution 10$^{4}$ times, each time producing a companion mass ratio distribution that represent the Galactic field distribution.  For every mass ratio distribution that we produced, we sampled it 10$^{6}$ times, and then performed a two sample Kolmogorov–Smirnov test (KS test) against the observed mass ratios of the binaries in the ONC with separations $>$ 20 AU.  The mean p-value resulting from this process is 0.02, likely ruling out that the observed ONC population and the Galactic field CMRD come from the same parent distribution, i.e. the observed mass ratio distribution in the ONC is statistically inconsistent with being drawn from the field distribution.  We also tested the observed ONC population against a flat CMRD ($\beta$ = 0.0) in the same manner as above, producing populations of 10$^{6}$ binaries 10$^{4}$ times.  This results in a mean p-value of 0.42, meaning we cannot rule out that the observed ONC population is drawn from the same parent population as the Galactic field CMRD.

\subsection{Comparison to Other Clusters} \label{subsec:clustercomparison}
The Galactic field is an amalgamation of all the stars, brown dwarfs, and planetary mass objects that are produced within SFRs over the lifetime of the Galaxy.  While the original environment of individual field objects is unknown, the binary populations within SFRs and older star clusters provide important information about the effects of environment and age on binary formation and evolution.

There have been many binary surveys of nearby SFRs and star clusters targeting brown dwarf primaries.  In the following comparisons, we focus on four SFRs and star clusters of various sizes and ages: Taurus-Auriga (Taurus), Chamaeleon I (Cha I), Upper Scorpius (Upper Sco), and the Pleiades.  To compare their VLMB populations to that of the ONC, we require that each primary have a mass $\leq$ 0.1 M$_{\odot}$ ($\geq$ M6 at 1 Myr) and that the survey is sensitive to binary systems of q $\geq$ 0.5 at projected separations $\geq$ 20 AU, except where specified.

In Taurus, a $\sim$ 1 Myr old low-mass stellar association, we combine the results of four surveys: \citet{Kraus2006, Konopacky2007, KrausHill2012, Todorov2014}.  Restricting their sample to members $\geq$ M6, there are two detections out of a total sample of 60 objects, a binary frequency of 3.3$_{-1.1\%}^{+4.1\%}$.  In a similar manner to the comparison of the field to the ONC, we sampled a binomial distribution that describes the binary frequency of Taurus and again integrated the ONC binomial distribution from 0 to the binary frequency of each iteration of the sampling procedure.  Each integration returns the probability that the observed Taurus binary frequency is consistent with our results in the ONC.  The p-value is 0.05, and therefore we do not find any strong evidence for an excess of very low mass binaries in the ONC compared to Taurus over projected separations 20 - 200 AU and mass ratios 0.5 - 1.0.  

In Cha I, a $\sim$ 2 Myr \citep{Luhman2004} low-mass stellar association, we combine the results of \citet{Ahmic2007}, \citet{Lafreniere2008}, and \citet{Todorov2014}.  Restricting their sample to members $\geq$ M6, there are zero detections out of a total sample of 24 sources.  We use the same procedure described above to calculate the probability that this binary frequency is consistent with our results in the ONC.  We obtain a p-value of 0.05, and therefore we find no strong evidence for an excess of VLMB in the ONC compared to Cha I over separations and mass ratios sampled. Additionally, \citet{Luhman2004} made a serendipitous detection of a 240 AU separation and q = 0.5 brown dwarf binary in Cha I, a potential indicator of the existence of wide brown dwarf binaries in this particular stellar association, analogous to our findings in the ONC.

Upper Sco is the nearest OB association that likely contains over 2000 stars \citep{Preibisch2008} and at 5 Myr old, could be an example of the future structure of the ONC, i.e. unbound, less dense and expanding outward \citep[cf.][]{HillenbrandHartmann}.  Combining the results of \citet{Biller2011} and \citet{KrausHill2012}, there is one detection (projected separation of 21 AU) out of a total sample of 40 targets.  Repeating the method from above, we calculate a p-value of 0.06 that the Upper Sco binary frequency is consistent with our results for the ONC over separations and mass ratios sampled.  If the ONC is unbound and evolves in a manner that resembles Upper Sco, then the brown dwarf binary population of the ONC does not need to be disrupted through additional processing mechanisms, but may still lose some binaries as the cluster dissolves into the field, resulting in more statistically similar populations.  It is expected that Upper Sco will directly contribute to the field as it expands.

Lastly, we compare the brown dwarf binary population of the ONC to the Pleiades, a 125 Myr old open cluster which likely contains over 1000 stars \citep{Adams2001}.  An ONC-like cluster has been suggested as a potential precursor to the Pleiades through N-body simulations \citep{Kroupa2001}.  Combining the samples of \citet{Martin2000} and \citet{Garcia2015}, there are zero detections out of a sample of 45 targets.  Importantly, \citet{Martin2000} are sensitive to companions down to 27 AU in projected separation, and therefore we must constrain the ONC sample to the same projected separation.  Over projected separations 27 - 200 AU and mass ratios 0.5 - 1.0, the ONC has a binary frequency of 10$_{-2.8\%}^{+5.8\%}$.  Following the method from above, we calculate a p-value of 0.02 that the binary frequency of the Pleiades is consistent with the binary frequency of the ONC population over separations and mass ratios sampled.  It is likely that the ONC will require further processing for its binary population to resemble that of the Pleiades at a comparable age.  However, larger sample sizes in both clusters will help determine the level of similarity between these binary populations.

\subsection{Implications} \label{subsec:implications}
Our analysis of the ONC VLMB population reveals an excess of wide (a = 23 - 230 AU) binaries relative to the Galactic field, over q = 0.5 - 1.0.  In comparison to other clusters, we find no strong evidence for an excess of VLMBs in the ONC relative to low mass stellar associations (e.g. Taurus and Cha I), but find a potential excess relative to the Pleiades.

Dynamical interactions have been invoked to explain how the binary frequency could decrease as unbound clusters dissolve into the Galactic field.  It has been suggested that the center of the ONC has undergone the cool-collapse process, i.e. contraction from a lower density state to a higher density state followed by an expansion back to a lower density state \citep[e.g.][]{Allison2009,Allison2010,ParkerGoodwinAllison2011}, where it currently remains expanding \citep{JonesWalker1988,dario2017}.  If the ONC is unbound and all its stellar components become a part of the Galactic field population, dynamical interactions may reconcile the discrepancy we observe between their very low mass binary populations.  On the other hand, if the ONC remains bound, it could ultimately resemble an open star cluster like the Pleiades \citep[as suggested by][]{Kroupa_Aarseth_Hurley2001}, and still most likely require further processing.

\citet{Weinberg1987} explored how effectively encounters with other stars and giant molecular clouds (GMCs) can disrupt wide binaries.  Using these calculations, \citet{Burg2003} determined that neither stellar nor GMC encounters in the Galactic field are likely to break up brown dwarf binaries with separations $<$ 10$^{4}$ AU, highlighting the potential need for encounters with other cluster members.  

We can calculate the typical lifetime of a specific binary within an ONC-like environment from eq. 28 of \citet{Weinberg1987}.  First, we approximate the stellar density in the region of the ONC in which our VLMBs exist, between 0.3 - 0.9 pc from the cluster center.  \citet{DaRio2014} measured the stellar mass density distribution of the ONC (their eq. 3) as follows: 

\begin{equation}
    \rho_{stars}(r) = 70 M_{Sun} pc^{-3} (\frac{r}{1 pc})^{-2.2}, r < 3 pc
\label{stellar_density}
\end{equation}

We then approximate the number density in this region by dividing the stellar mass density by the characteristic mass of 0.3 M$_{\odot}$ \citep{DaRio2012}.  For a VLMB of total mass 0.1 M$_{\odot}$ and separation of 100 AU at 0.3 pc, the approximate stellar number density is 3.7 x 10$^{3}$ stars pc$^{-3}$ and the characteristic lifetime is 5 Myr.  The mean age of the ONC is estimated to be 2.2 Myr \citep{Reggiani2011}, so these observed VLMBs may not have existed long enough to be disrupted but they may not ultimately survive the current dense environment of the ONC.  Importantly, \citet{Reggiani2011} and \citet{Beccari2017} find an age spread in the ONC stellar population.  This leaves open the possibility that these wide binaries formed after the cool collapse phase and did not encounter an environment of significantly higher density that would almost assuredly destroy such low binding energy binaries.  Instead, the disruptions may occur over the next several Myr and could explain both the discrepancy in the binary frequency and the CMRD of the ONC relative to the Galactic field.

This scenario is consistent with that of \citet{Kroupa2003} who shows that dynamical interactions of wide ($>$ 20 AU) VLMBs in an ONC-like cluster can happen in $\sim$ 1 Myr.  \citet{ParkerGoodwin2011} also show that clusters represented by virialized Plummer spheres of initial density comparable to the current density in the ONC \citep[half-mass radius = 0.8 pc, see][]{mccaughrean_Stauffer1994,HillenbrandHartmann} can disrupt some wide VLMBs at separations $>$ 10 AU, although most of these binaries remain unaffected.  While the ONC may already be significantly dynamically mixed, these simulations leave open the possibility for further dynamical interactions of VLMBs.

Conversely, we considered the possibility that these wide binaries will be hardened to the smaller separations seen in the Galactic field instead of disrupted through dynamical interactions.  Many simulations of star clusters that account for dynamical interactions show that wider binaries are preferentially destroyed due to their low binding energy. \citet{KroupaBurkert2001,Kroupa_Aarseth_Hurley2001, ParkerGoodwinAllison2011, ParkerGoodwin2012} all arrive at a decrease in wide binaries with no subsequent increase in small separation binaries after dynamical processing, indicative of the fact that wide binaries are not likely to harden.

Importantly, an excess of VLMB is not inconsistent with our results from Paper I.  We found that the low-mass stellar (0.1 M$_{\odot}$ $<$ M$_{prim}$ $<$ 0.6 M$_{\odot}$) binary frequency was consistent with the Galactic field, leaving no need for further dynamical interactions.  Although these stellar binaries have similar separations to the VLMBs, the total binding energy of stellar binaries is significantly higher than that of the VLMBs.  Additionally, the characteristic lifetime of the stellar binaries is between 4-20 times longer than that calculated for the VLMBs, using the \citet{Weinberg1987} derivation.  If the ONC continues to expand and eventually become a part of the Galactic field, the stellar binaries are very unlikely to be disrupted while the VLMBs are significantly more likely to be disrupted before large changes in the density.

On the other hand, \citet{ParkerGoodwin2012} stress the potential of stochasticity in the destruction of binaries within ONC-like star clusters.  They show that over separations 62 to 620 AU the number of stellar binaries destroyed after 1 Myr can vary by a factor of $>$ 2, with a statistically significant difference, after only 10 realizations of the same cluster.  Any apparent statistical difference of an observed young binary population of one star cluster could be a result of stochasticity in the destruction of binaries rather than evidence for a distinct formation mechanism or environmental impact.

Future observations with the \textit{James Webb Space Telescope} could easily identify many brown dwarfs in the ONC down to 1 M$_{Jup}$ from which a binary survey of separations $>$ 20 AU can be completed.  A multi-filter photometric program (e.g. GTO program 1256, PI: M. McCaughrean) devoted to observing low mass cluster members would greatly increase the sample size in the ONC (on the order of 100s) and be able to identify companions at significantly lower mass ratios.  This type of program is also planned for NGC 2024 (GTO program 1190, PI: M. Meyer), and they will serve as a proof of concept to be carried forward into other young clusters.  In the further future, the application of extremely large telescopes (ELTs) in star clusters would uncover the binary population down to a few AU, overlapping the separation range in which we find most field VLMBs for a more thorough comparison of the VLMB population of young, star clusters.   Importantly, ELTs with adaptive optics will have to target individual objects with smaller fields of view than available with wide-field space based imaging.  These two strategies in tandem will be able to identify a vast majority of binaries over multiple orders of magnitude in separation.

\section{Conclusion} \label{sec:conclusion}
We conducted the first brown dwarf binary survey of the ONC sensitive to projected separations of 10 AU using an improved version of our double-PSF fitting technique from Paper I.  This technique is applicable to much of the available HST archival data due to the PSF stability from space, as long as empirical PSF libraries are available.  To summarize the results of our survey:

1)  Our binary detection routine is capable of identifying companions at 0.025" for high S/N sources, below the diffraction limit of HST.  This resolution limit is competitive with that attained through the aperture masking interferometry technique with ground-based adaptive optics on 8m telescopes, e.g. \citet{Duchene2018}, and kernel phase interferometry with HST, e.g. \citet{Pope2013}.

\vspace{0.35cm}
2)  We detected 7 candidate very low mass binaries, 6 of which were found in multiple filters, from our total sample of 75 very low-mass star and brown dwarf primaries (M$_{prim}$ $\leq$ 0.1M$_{\odot}$), two of which were previously found by \citet{Reipurth2007}.  The closest companion detection is at a projected separation of 11 AU.

\vspace{0.35cm}
3)  For our sample of M$_{prim}$ $\leq$ 0.1M$_{\odot}$, over physical separations of 23 - 230 AU and mass ratios of 0.5 - 1.0, we observed a binary frequency of 12$_{-3.2\%}^{+6.0\%}$ in the ONC.  This result is distinct from the Galactic field over the same sensitivity range, 0.9\%$\pm 0.6$\%, with a probability of the field binary frequency being the true ONC binary frequency equal to 10$^{-6}$.

\vspace{0.35cm}
4) Based on the current high density environment of the ONC, dynamical interactions over the next few Myr could destroy VLMBs with low binding energies (e.g. lower mass ratios).  If this is an important process within ONC-like clusters \citep[as suggested by][]{Weinberg1987,Burg2003}, dynamical interactions could sculpt the VLMB population of the ONC to resemble that of the Galactic field or the Pleiades.

\begin{acknowledgments}
We thank an anonymous referee for a prompt and helpful report which improved the presentation and the content of this manuscript.  We would like to thank Jay Anderson for many productive discussions on PSF modeling and the implementation of his PSF code, as well as Megan Kiminki for contributions to the construction of our code.  Megan Reiter received funding from the European Unions Horizon 2020 research and innovation programme under the Marie Skodoska-Curie grant agreement No. 665593 awarded to the Science and Technology Facilities Council.  This work is based on observations made with the NASA/ESA Hubble Space Telescope, obtained from the data archive at the Space Telescope Science Institute. STScI is operated by the Association of Universities for Research in Astronomy, Inc. under NASA contract NAS 5-26555. Support for this work was provided by NASA through grant number HST-AR-15047.001-A from the Space Telescope Science Institute, which is operated by AURA, Inc., under NASA contract NAS 5-26555.
\end{acknowledgments}

\bibliography{sample63}{}
\bibliographystyle{aasjournal}

\appendix
\section{Singles}
\begin{deluxetable}{ccc}[b]
\tablenum{3}
\tablecaption{This table contains the sources in our sample for which we did not find a companion.  Of our 75 sources, 67 did not have companions and are labeled as single sources over the sensitivity ranges in which our code could identify a companion.  We list the adopted spectral type taken from the spectroscopic studies described in Sec. \ref{subsec:subsample}}
\tablewidth{0pt}
\tablehead{
\colhead{RA (J2000)}  & \colhead{Dec (J2000)}  & \colhead{Adopted Spectral Type} }
\decimalcolnumbers
\startdata
83.57221 & -5.37997 & M7.5\\
83.59996 & -5.47631 & M6 \\
83.66675 & -5.43442 & M6 \\
83.67433 & -5.36375 & M7.5 \\
83.67488 & -5.35889 & M6 \\
83.70900 & -5.33339 & M6 \\
83.73292 & -5.35236 & M6.25 \\
83.73729 & -5.35608 & M6.75 \\
83.74283 & -5.42589 & M6 \\
83.75375 & -5.35203 & M8 \\
83.75542 & -5.38500 & M9 \\
83.75967 & -5.51961 & M7 \\
83.76292 & -5.42333 & M7.5 \\
83.76546 & -5.37933 & M6.5 \\
83.76746 & -5.33667 & M7.75 \\
83.76858 & -5.42403 & M8.5 \\
83.76950 & -5.43078 & M8.5 \\
83.77275 & -5.37511 & M6 \\
83.77417 & -5.38775 & M6.5 \\
83.77688 & -5.38753 & M6 \\
83.77942 & -5.41692 & M8 \\
83.78013 & -5.44406 & M6.5 \\
83.78379 & -5.37708 & M8 \\
83.78425 & -5.38144 & M7.5 \\
83.78463 & -5.32700 & M9 \\
83.78475 & -5.38942 & M8 \\
83.78517 & -5.38469 & M7 \\
83.78608 & -5.33956 & M9.5 \\
83.78763 & -5.39064 & M6 \\
83.78833 & -5.43486 & M8 \\
\enddata
\end{deluxetable}

\begin{deluxetable}{ccc}[b]
\tablewidth{0pt}
\startdata
83.78833 & -5.42556 & M7.5 \\
83.79079 & -5.40186 & M6.5 \\
83.79179 & -5.41708 & M9 \\
83.79250 & -5.35006 & M7.5 \\
83.79325 & -5.41433 & M7.5 \\
83.79458 & -5.41472 & M8 \\
83.79521 & -5.43247 & M6.5 \\
83.79646 & -5.41014 & M9 \\
83.81150 & -5.47550 & M8.5 \\
83.81671 & -5.36475 & M8 \\
83.81883 & -5.44289 & M8.75 \\
83.82521 & -5.36144 & M7.5 \\
83.82533 & -5.38383 & M9 \\
83.82625 & -5.45806 & M8.75 \\
83.82763 & -5.44206 & M8 \\
83.83004 & -5.51750 & M6 \\
83.83600 & -5.37811 & M8 \\
83.83708 & -5.42625 & M8 \\
83.83758 & -5.38175 & M7 \\
83.83896 & -5.42639 & M8 \\
83.84038 & -5.44794 & M7.75 \\
83.84196 & -5.46956 & M6 \\
83.84217 & -5.41878 & M8 \\
83.84633 & -5.22872 & M6.5 \\
83.84671 & -5.47606 & M7.75 \\
83.84788 & -5.25647 & M6.5 \\
83.84808 & -5.39750 & M9 \\
83.85429 & -5.41067 & M7 \\
83.85529 & -5.40181 & M7 \\
83.85688 & -5.30117 & M6.75 \\
83.86021 & -5.50456 & M6 \\
83.86546 & -5.48094 & M7 \\
83.87292 & -5.42700 & M6 \\
83.88200 & -5.50114 & M6.5 \\
83.88308 & -5.52989 & M8 \\
83.88683 & -5.30550 & M7.5 \\
83.89317 & -5.44989 & M7 
\enddata
\end{deluxetable}

\end{document}